\documentclass{optica-article-modif}

\journal{opticajournal}

\articletype{Research Article}

\usepackage{lineno}

\usepackage{graphicx}
\usepackage{amsfonts}
\usepackage{bm}
\usepackage{hyperref,url}

\usepackage{xcolor}
\usepackage{soul}
\usepackage{siunitx}
\usepackage{todonotes}
\usepackage{ulem}

\newcommand{\bd}{\begin{displaymath}}
\newcommand{\ed}{\end{displaymath}}
\newcommand{\be}{\begin{equation}}
\newcommand{\ee}{\end{equation}}
\newcommand{\bs}{\begin{subequations}}
\newcommand{\es}{\end{subequations}}
\newcommand{\ba}{\begin{eqnarray}}
\newcommand{\ea}{\end{eqnarray}}

\begin{document}

\title{Exploring the dynamics of finite-energy Airy beams: A trajectory analysis perspective}

\author{\'Angel S. Sanz\authormark{*} and Rosario Mart{\'\i}nez-Herrero,\authormark{**}}

\address{
Department of Optics, Faculty of Physical Sciences, Universidad Complutense de Madrid, Pza.\ Ciencias 1, 28040 Madrid, Spain
}

\email{\authormark{*}a.s.sanz@fis.ucm.es; \authormark{**}r.m-h@fis.ucm.es}




\begin{abstract}
In practice, Airy beams can only be reproduced in an approximate manner, with
a limited spatial extension and hence a finite energy content.
To this end, different procedures have been reported in the literature, based
on a convenient tuning of the transmission properties of aperture functions.
In order to investigate the effects generated by the truncation and hence the propagation
properties displayed by the designed beams, here we resort to a new perspective based on
a trajectory methodology, complementary to the density plots more commonly used to study
the intensity distribution propagation.
We consider three different aperture functions, which are convoluted with an ideal Airy beam.
As it is shown, the corresponding trajectories reveals a deeper physical insight about the
propagation dynamics exhibited by the beams analyzed due to their direct connection with
the local phase variations undergone by the beams, which is in contrast with the global
information provided by the usual standard tools.
Furthermore, we introduce a new parameter, namely, the escape rate, which allow us to
perform piecewise analyses of the intensity distribution
without producing any change on it, e.g., determining unambiguously how much energy flux
contributes to the leading maximum at each stage of the propagation, or for how long
self-accelerating transverse propagation survives.
The analysis presented in this work thus provides an insight into the behavior of
finite-energy Airy beams, and therefore is expected to contribute to the design and
applications exploiting this singular type of beams.
\end{abstract}


\section{Introduction}
\label{sec1}

In 1979, Berry and Balazs found \cite{berry:AJP:1979} that the free-propagation Schr\"odinger
equation, a rather simple and even trivial equation, admitted a rather counter-intuitive
nondispersive solution.
Provided the initial ansatz is described by an Airy function, these authors proved that the
time-evolution of the corresponding probability density remains shape-invariant during the
whole propagation.
At the same time, without the presence of any external potential, these wave packets undergo
a self-accelerating motion pushing them forwards.
When this behavior is analyzed within the framework of the equivalence principle, as it was
done by Greenberger \cite{greenberger:AJP:1980}, it can be shown that Schr\"odinger's equation
can be recast as the time-independent equation describing a free falling mass, of which the
eigensolutions correspond precisely to Airy functions.
Later on, Unnikrishnan and Rau \cite{unnikrishnan:AJP:1996} proved the uniqueness of these
solutions to Schr\"odinger's equation as the only free-space self-accelerating nondispersive one.

The beams conjectured by Berry and Balazs have not become a reality until their experimental
realization, formerly produced in 2007 by Siviloglou {\it et al.}\ \cite{christodoulides:PRL:2007},
taking advantage of the isomorphism between the free-space Schr\"odinger equation for a
non-relativist particle with a mass $m$, and the paraxial Helmholtz equation for monochromatic
light propagating in a linear medium with a refractive index $n$.
Ever since, Airy beams have become one of the relevant components in the puzzle of structured
light \cite{forbes:NatPhotonics:2021,wang:FrontPhys:2021}, a very active and promising research
field strongly developed over the last decade due
to the interesting fundamental and technological applications brought in by conveniently
designed light beams.
In this regard, different control mechanisms for their spatial manipulation \cite{Zhang:ApplSci:2017}
and applications \cite{christodoulides:Optica:2019} have been discussed in the literature.

From the point of view of their experimental implementation, the production of Airy beams
introduces an important technical challenge: how is it possible to reproduce such beams,
with their characteristic features, if experimental beams have a limited spatial extension,
and hence the infinite tail of ideal Airy beams becomes unaffordable?
This is the same to say that, although the energy content of an ideal Airy beam is infinite,
in practice their energy content will be finite due to the unavoidable spatial truncation
under experimental conditions.
The question, therefore, is how to introduce such truncations in order to warrant the
preservation of the Airy beam properties (self-accelerated transverse displacement and
shape invariance of the intensity distribution) for relatively large propagation distances
with respect to the input plane at which they are originated.

A way to overcome this drawback was formerly proposed by Siviloglou and Christodoulides
\cite{christodoulides:OptLett:2007}, which consisted in using an exponential aperture, such that
it has no influence on the beam at the front of the beam (the front part of the Airy
function cancels out the increasing exponential values), while its tail of secondary
maxima is gradually annihilated in the rear part (by the quickly decreasing of the
exponential function).
As it was shown both theoretically and experimentally \cite{christodoulides:PRL:2007}, this
method allows a propagation of the beam for rather long distances, along which its
distinctive traits are preserved.
However, it does not prevent the eventual appearance of diffraction and, hence, the
irreversible loss of such features.
This is an inconvenience if the method is going to be considered as a resource to keep the
Airy beam properties for longer distances, not only in linear media, but also in the
nonlinear domain \cite{Zhang:ApplSci:2017}.

Because of its practical implications, as seen above, here we tackle the study of how
the functional form of the aperture function affects the properties displayed by the
truncated Airy beam that it produces.
However, unlike the procedure followed in \cite{christodoulides:OptLett:2007}, instead of
directly introducing the truncation in positions, we proceed through the Fourier space,
Since the action of the aperture function on the ideal Airy beam is the convolution of the
latter with the former, the Fourier transform of the truncated Airy beam corresponds to
the direct product of the Fourier transforms of the aperture function and the ideal Airy
beam.
A similar procedure has been considered in the literature to investigate different
properties of finite-energy Airy beams, for instance, the periodic inversion and phase
transition in non-homogeneous media \cite{belic:OE:2015}, their manipulation in dynamic
parabolic potentials \cite{belic:AnnPhysik:2020}, or the achievement of off-axis
auto-focusing and transverse self-acceleration \cite{Xu:OE:2022}.

Furthermore, in our case, to investigate the properties of these aperture functions,
we introduce a trajectory-based methodology, which allows us to follow the transverse energy
flux at a local level, in a hydrodynamic manner \cite{sanz:JOSAA:2022}.
The two key elements in this methodology are, on the one hand, the effective transverse
velocity field associated with the variations undergone by the transverse intensity
distribution during the propagation.
On the other hand, by integrating such a velocity field with different input or initial
conditions, we obtain swarms of trajectories that provide us with valuable local information
of how such variations take place.
Despite of providing us with complementary information, it is worth highlighting that both
elements are directly connected to the local spatial variations undergone by the beam phase
during propagation.
Thus, in the ideal case, the field will be linear and the trajectories will be parabolas,
while in the case of truncated beams an additional contribution is shown to appear in the
velocity field, which disturbs the typical parabolic evolution of the trajectories.
Apart from this kind of dynamics information, we also show how the trajectories make
possible to determine how much a given feature degrades as the beam propagates forward,
like the main maximum of a finite-energy Airy beam, in a very precise manner.
In this work, we thus merge the information obtained from global behaviors, studied by means
of the standard propagation of the intensity distribution, with local phase effects, which
are examined by virtue of the computation and subsequent analysis of the corresponding flux
trajectories.

The organization of this work is as follows.
In Sec.~\ref{sec2}, we introduce the general theoretical aspects of the methodology used
here as well as the specific details in the particular application to aperture functions.
Moreover, we also discuss the main details of the aperture functions considered in the
analysis.
In Sec.~\ref{sec3}, we present and discuss the mains results obtained for the three aperture
functions considered.
The simulations presented include both propagation properties and also information extracted
from trajectory statistics.
Finally, Sec.~\ref{sec4} summarizes the main conclusions from this work.


\section{Theory}
\label{sec2}


\subsection{General aspects}
\label{sec21}

Consider the paraxial monochromatic scalar field
\begin{equation}
 \Psi({\bf r}_\perp,z) = \psi({\bf r}_\perp,z) e^{ikz} ,
 \label{eq1}
\end{equation}
where ${\bf r}_\perp =(x,y)$ and $k = 2\pi n/\lambda_0$ (with $\lambda_0 = \omega/c$).
The complex-valued amplitude $\psi({\bf r}_\perp,z)$ satisfies the paraxial Helmholtz
equation,
\begin{equation}
 i\ \frac{\partial \psi({\bf r}_\perp,z)}{\partial z}
 = - \frac{1}{2k}\ \nabla_\perp^2 \psi({\bf r}_\perp,z) ,
 \label{eq2}
\end{equation}
where $\nabla_\perp^2 = \partial^2/\partial x^2 + \partial^2/\partial y^2$ is the
transverse Laplacian.
If this equation is multiplied on both sides by $\psi^*({\bf r}_\perp,z)$, and then we
subtract to the resulting equation its complex conjugate version, we obtain
\begin{equation}
 \frac{\partial |\psi|^2}{\partial z}
  = - \frac{1}{2ik} \left[ \psi^* \left( \nabla_\perp^2 \psi \right)
    - \left( \nabla_\perp^2 \psi^* \right) \psi \right] .
 \label{eq3}
\end{equation}
Rearranging terms on the right-hand side of this equation, it can be further simplified in
the form of the following transport equation
\begin{equation}
 \frac{\partial |\psi|^2}{\partial z}
  = - \nabla_\perp \cdot \left[ |\psi|^2
     \frac{1}{k}\ {\rm Re} \left( \frac{-i\nabla_\perp \psi}{\psi} \right) \right] ,
 \label{eq4}
\end{equation}
which describes the transversal changes undergone by the intensity distribution,
$|\psi|^2$, as it propagates along the $z$-direction.
Such changes are determined by the term on the right-hand side, where the flux density
\cite{margenau-moseley-bk} is associated here with a transverse electromagnetic flux
vector,
\begin{eqnarray}
 {\bf j}_\perp ({\bf r}_\perp,z) & = & \frac{1}{2ik}
    \left\{ \psi^*({\bf r}_\perp,z) \left[ \nabla_\perp \psi({\bf r}_\perp,z) \right]
   - \left[ \nabla_\perp \psi^*({\bf r}_\perp,z) \right] \psi({\bf r}_\perp,z) \right\}
 \nonumber \\
 & = & |\psi({\bf r}_\perp,z)|^2 {\bf v}_\perp ({\bf r}_\perp,z) .
 \label{eq5}
\end{eqnarray}
Actually, according to the usual definition of a flux density, the second line of
Eq.~\eqref{eq5} is specified in terms of the intensity distribution and the transport
field
\begin{equation}
 {\bf v}_\perp ({\bf r}_\perp,z) =
  \frac{1}{k}\ {\rm Re} \left( \frac{-i\nabla_\perp \psi}{\psi} \right) .
 \label{eq6}
\end{equation}
This field quantifies the local rate of change of the flux, analogous to a local velocity
field describing the flow of energy on a given point ${\bf r}_\perp$ at a given value of
the $z$ distance.
Hence, from now on, we shall refer to ${\bf v}_\perp$ as the effective transverse velocity
field.

It is worth noting that, although here we have proceeded directly from the paraxial Helmholtz equation, Eq.~\eqref{eq2}, in order to reach the expression for the flux
\eqref{eq5}, in analogy to how we could have proceeded in quantum mechanics with
Schr\"odinger’s equation to obtain the quantum flux \cite{sanz-bk-1}, it could have also
be done considering the full vector nature of the electromagnetic field and then assuming
paraxiality conditions, as formerly done by Allen {\it et al.} \cite{allen:PRA:1992}.
In this latter case, one explicitly observes that the time-averaged Poynting vector consists
of two contributions, one simply related to the forward propagation in the paraxial case,
and another one dealing with the space variations undergone by the field amplitude, which,
in turn, are connected to the field local phase variations.
This convenient decoupling between the longitudinal and the transverse contributions in the
Poynting vector has been considered more recently, for instance, by Broky {\it et al.}
\cite{broky:OE:2008} to precisely investigate the self-healing properties of Airy beams
in two dimensions.

As it can be noticed, if the complex field amplitude is recast in the following polar form
\begin{equation}
 \psi({\bf r}_\perp,z) = A({\bf r}_\perp,z) e^{iS({\bf r}_\perp,z)} ,
 \label{eq7}
\end{equation}
where both $A({\bf r}_\perp,z)$ and $S({\bf r}_\perp,z)$ are real-valued fields, the above
defined velocity field reads as
\begin{equation}
 {\bf v}_\perp ({\bf r}_\perp,z) =
  \frac{1}{k}\ \nabla_\perp S({\bf r}_\perp,z) .
 \label{eq8}
\end{equation}
This expression unambiguously shows that any change in the transverse flux is directly
associated with the local (transverse) spatial variations undergone by the phase of the
field phase during its propagation, which eventually leads to measurable changes in the
intensity distribution.
Accordingly, to better understand these physical consequences, one can compute ensembles of
trajectories obtained from the integration of the equation of motion
\begin{equation}
 \frac{d{\bf r}_\perp (z)}{dz} = {\bf v}_\perp ({\bf r}_\perp,z)
  = \frac{{\bf j}_\perp ({\bf r}_\perp,z)}{|\psi ({\bf r}_\perp,z)|^2} ,
 \label{eq9}
\end{equation}
which will provide us with first-hand local information directly associated with the phase of
the field (or, more strictly speaking, its spatial phase variations) rather than from the
profile displayed by the intensity distribution (from which any phase effect is indirectly
inferred a posteriori).
In this regard, these trajectories have already been used to describe light propagation
through optical fibers \cite{sanz:JOSAA:2012}, to account for the differences between
Gaussian beams and Maxwell-Gauss beams \cite{sanz:ApplSci:2020}, or, more recently, to
describe the dynamics of ideal and finite-energy Airy beams \cite{sanz:JOSAA:2022} or to
investigate light focusing \cite{sanz2023quantum}.
Furthermore, it has also been used to establish a connection between quantum and optical
analogs given the isomorphism between the paraxial Helmholtz equation and the time-dependent
Schr\"odinger equation \cite{sanz-bk-1}.

In analogy to their quantum-mechanical counterparts \cite{sanz-bk-1}, the trajectories
obtained after integrating Eq.~\eqref{eq9} have three distinctive properties with practical
interest concerning the problem that we are dealing with here:
\begin{itemize}
 \item[(i)] The trajectories cannot get across the same point ${\bf r}_\perp$ at the same
 value of $z$.
 This is a direct consequence that arises from the single valuedness of the solutions of
 the paraxial Helmholtz equation \eqref{eq2}, except for a constant phase factor.
 From the solution in polar form \eqref{eq7}, we note that this phase factor corresponds
 to an integer multiple of $2\pi$, which does not affect the trajectory dynamics, as it
 can be inferred from the Eq.~\eqref{eq8}.

 \item[(ii)] Consider a closed curve $\mathcal{C}({\bf r}_\perp,0)$ at $z=0$ defined by
 a set of initial conditions ${\bf r}_\perp(0)$ for Eq.~\eqref{eq9}.
 At any other value of $z$, $\mathcal{C}({\bf r}_\perp,0)$ will uniquely evolve into another
 closed curve $\mathcal{C}({\bf r}_\perp,z)$ consisting of the positions ${\bf r}_\perp(z)$
 of the trajectories started at ${\bf r}_\perp(0)$.

 \item[(iii)] Following properties (i) and (ii), the partial energy content within the
 surface $\Sigma({\bf r}_\perp,z)$ determined by the curve $\mathcal{C}({\bf r}_\perp,z)$,
 defined as
 \begin{equation}
  E_{\Sigma({\bf r}_\perp,z)} =
   \int_{\Sigma({\bf r}_\perp,z)} |\psi({\bf r}_\perp,z)|^2 d{\bf r}_\perp
  \propto \lim_{N\to\infty} \sum_{i=1}^N
  \delta \left[ \{{\bf r}_\perp \in \Sigma({\bf r}_\perp,z)\} - {\bf r}_\perp(z) \right] ,
  \label{eq10}
 \end{equation}
 remains constant regardless of the value of $z$ and the shape acquired by
 $\mathcal{C}({\bf r}_\perp,z)$ along its propagation \cite{sanz:AnnPhys:2013}.
 Note, on the right-hand side of the second equality, that $E_{\Sigma({\bf r}_\perp,z)}$ is
 directly related to the set of all trajectories laying within $\mathcal{C}({\bf r}_\perp,z)$
 and with initially distributed within $\mathcal{C}({\bf r}_\perp,0)$ according to
 $|\psi({\bf r}_\perp,0)|^2$.
\end{itemize}
Specifically, as it is shown below, here we use these properties to quantitatively analyze
some aspects associated with the leading peak of the input finite-energy Airy beams.


\subsection{Finite-energy Airy beams}
\label{sec22}

Let us consider the particular case of the one-dimensional Helmholtz equation, although
this will be done, for simplicity, in terms of the reduced coordinates $\tilde{x} = x/x_0$
($x_0$ denotes an arbitrary typical value for the transverse coordinate) and
$\tilde{z} = z/k x_0^2 = \lambda_0 z/2\pi n x_0^2$ \cite{sanz-bk-1},
\begin{equation}
 i\ \frac{\partial \psi(x,z)}{\partial z} = - \frac{1}{2} \frac{\partial^2 \psi(x,z)}{\partial x^2}
 \label{eq11}
\end{equation}
(to simplify notiation, from now on tildes will be removed).
As Berry and Balazs showed \cite{berry:AJP:1979}, for wave packets with an initial amplitude described
by an Airy function, propagating according to Eq.~\eqref{eq11},
their amplitude remains shape invariant and undergoes a transverse displacement that increases rapidly with $z^2$:
\begin{equation}
 \psi(x,z) = e^{i(x-z^2/6) z/2} Ai(x - z^2/4) .
 \label{eq12}
\end{equation}
Alternatively, the propagation of these beams can be described at a local level in terms of
trajectories associated with the transverse energy flux \cite{sanz:ApplSci:2020},
\begin{equation}
 j(x,z) = {\rm Im} \left[ \psi^*(x,z) \frac{\partial \psi(x,z)}{\partial x} \right] .
 \label{eq13}
\end{equation}
Specifically, the trajectories arise from the integration along $z$ of the equation
of ``motion''
\begin{equation}
 \frac{dx(z)}{dz} = v(x,z) = \frac{j(x,z)}{I(x,z)} ,
 \label{eq14}
\end{equation}
where $v(x,z)$ is the local field describing the energy flow
along the transverse direction ($x$) at a given value of the longitudinal coordinate ($z$),
and $I(x,z) = |\psi(x,z)|^2 = |Ai(x -z^2/4)|^2$ is the beam intensity distribution.

Because the field \eqref{eq14} is directly related to the beam local phase variations along the transverse direction, it becomes a suitable tool to explore any property
associated with the phase and, more specifically, with effects due to possible variations induced on it (see, for instance, Refs.~\cite{sanz:JOSAA:2022,sanz2023quantum}).
Note that for ideal Airy beams, $v(x,z)$ depends linearly on the $z$-coordinate and hence the
associated trajectories are fully analytical:
\begin{equation}
 x(z) = x(0) + \frac{z^2}{4} .
 \label{eq14b}
\end{equation}
These flux trajectories are parabolas, in agreement with the renowned parabolic transverse
acceleration undergone by the beam during its propagation along $z$.

In the case of non-ideal Airy beams, that is, realistic beams with a finite energy
content, because the infinite tail of the ideal Airy beam is not physically realizable
in practice, the corresponding truncation introduces extra phase factors that eventually
lead to the gradual dispersion of the beam.
In other words, the beam will gradually lose both its shape invariance and its transverse
self-accelerated propagation beyond a certain spatial range.
In this regard, the above trajectory-based approach constitutes a suitable tool to detect and
explore any change directly related to the phase at a local level.

The truncation can be introduced by directly acting on the tail of the beam, as it was
formerly proposed by Siviloglou and Christodoulides \cite{christodoulides:OptLett:2007},
who considered an exponentially decaying factor acting on an input (ideal) Airy beam, i.e.,
\begin{equation}
 \psi(x,0) = Ai(x) e^{\gamma x} .
 \label{eq16}
\end{equation}
The propagated beam in this case can be analytically determined and reads as
\begin{equation}
 \psi(x,z) = e^{i(x-z^2/6) z/2 + \gamma (y - z^2/4) - i\gamma^2 z/2} Ai(y) ,
 \label{eq17}
\end{equation}
where the complex argument $y = x - z^2/4 + i\gamma z$ in the Airy function makes the
latter to become also complex, thus leading to the appearance of another phase factor,
apart from those arising from the multiplying exponential prefactor.
More specifically, the expression for the velocity field becomes \cite{sanz:JOSAA:2022}
\begin{equation}
 v(x,z) = \frac{z}{2}
  + \frac{\partial}{\partial x} \left\{ {\rm arg} \left[ Ai(y) \right] \right\} ,
\label{eq18}
\end{equation}
which leads to a non-analytical equation of motion, where we clearly observe that the
extra term (second term on the right-hand side) is responsible for the deviation of
the trajectories (flux) from the typical parabolic shape associated with an ideal Airy
beam.

In the above case, the field can be represented as a product of the ideal Airy beam and
an exponential.
Hence, the corresponding transverse spectrum,
\begin{equation}
 \tilde{\psi}(k_x) = e^{ik_x^3/3 - \gamma (k_x - i \gamma/2)^2 + \gamma^3/12} ,
 \label{eq19}
\end{equation}
is the direct product of the transverse spectrum corresponding to the ideal Airy beam,
\begin{equation}
 \tilde{\psi}(k_x) = e^{ik_x^3/3} ,
 \label{eq20}
\end{equation}
and a complex Gaussian function.
Taking into account this result and that, according to the convolution theorem, the Fourier
transform of the convolution of two functions is the direct product of the Fourier transforms
of such functions, we proceed now as follows to investigate the effects induced by known
aperture functions on the flux dynamics.
Consider that the finite-energy Airy beam results from the convolution function
\begin{equation}
 \psi(x,z) = \int w(x') Ai(x - x',z) dx' ,
 \label{eq21}
\end{equation}
where $w(x)$ represents a given aperture function.
From a formal point of view, here we are going to assume that $w(x)$ satisfies the following
two properties: ($i$) it is $L^2$ in order to warrant the finiteness of the energy content of
$\psi(x)$, and ($ii$) under some control parameter it approaches a Dirac $\delta$-function,
which allows us to recover the original (ideal) infinite energy Airy beam.

Thus, according to the convolution theorem, the Fourier transform of \eqref{eq21} reads as
\begin{equation}
 \tilde{\psi}(k_x) = \tilde{w}(k_x) e^{ik_x^3/3} ,
 \label{eq22}
\end{equation}
from which we already infer that any additional phase effect is going to be associated with
the presence of the new term $\tilde{w}(k_x)$, as the Airy prefactor only governs the
curvature of the flux.
More importantly, in the current case, note that the transverse power spectrum of the ideal
Airy beam is totally flat, so the parabolic curvature of the flux only comes from the cubic
phase factor of the Fourier transform.
The appearance of $\tilde{w}(k_x)$ in \eqref{eq22} is not only going to affect the phase,
as seen in \eqref{eq18}, but it may also induce a modulation of the spectral components,
thus removing the flatness of the power spectrum.
Both factor are therefore responsible for the eventual diversion of the propagation of the
finite-energy beam from the one displayed by an ideal Airy beam.

Taking into account the functional form of a propagated Airy beam, Eq.~\eqref{eq11}, the
amplitude \eqref{eq21} can be recast as
\begin{equation}
	\psi(x,z) = e^{i(x - z^2/6)z/2} \mathcal{A}i(x,z) ,
	\label{eq23}
\end{equation}
where
\begin{equation}
 \mathcal{A}i(x,z) \equiv \int w(x') Ai(x - x' - z^2/4)\ e^{-izx'/2} dx' .
 \label{eq24}
\end{equation}
The amplitude \eqref{eq23} leads to a generalized equation of motion that now reads as
\begin{equation}
 v(x,z) = \frac{z}{2}
  + \frac{\partial}{\partial x} \left\{ {\rm arg} \left[ \mathcal{A}i(x,z) \right] \right\} ,
 \label{eq25}
\end{equation}
containing the action of an extra term on the right-hand side.
This extra term is directly connected with the spectrum of the aperture functions, which
induces a new local phase configuration apart from the truncation of the beam, thus leading
to the distortion of the beam with respect to its ideal counterpart.

So far the trajectories obtained after integrating Eq.~\eqref{eq25} provide us with an idea
of the newly induced phase effects on the flux dynamics.
However, we can also extract valuable quantitative information regarding the finite-energy
beams if we consider particular ensembles of such trajectories and the above properties
satisfied by the trajectories arising from Eq.~\eqref{eq25}.
Here, in particular, we consider two quantities to investigate in a more quantitative manner
the dynamics associated with the leading maximum of the input finite-energy Airy beans
generated:
\begin{itemize}
 \item[(i)] The escape rate, $N(z)$, defined as the partial energy content that remains
 confined within two parabolic trajectories delimiting the main maximum of an ideal Airy
 beam, given that the main maximum for the three input beams here considered basically
 overlaps with that of the ideal Airy beam.
 This quantity is computed according to Eq.~\eqref{eq10} and the $\mathcal{C}(x,0)$ is
 determined by the above mentioned limiting trajectories.
 In order to produce a fair sampling in this single-event dynamics, the initial conditions
 are randomly chosen according to the weight $|\psi(x,0)|^2$
 (see Appendix~\ref{app} for additional technical details).
 Depending on the functional form of the aperture function, the escape rate will fall faster
 or slower, thus indicating either a faster or a slower deviation with respect to the ideal
 Airy beam.

 \item[(ii)] The average position, $x_{\rm av}(z)$, determined at each $z$-value as the direct
 average taking into account all trajectories in the ensemble.
 Again, like the escape rate, this quantity supplies local information about how fast the
 main peak disappears, and hence how fast the finite-energy beam looses the properties that
 characterize an ideal Airy beam.
\end{itemize}


\subsection{Aperture functions}
\label{sec23}

Specifically, to explore the feasibility of aperture functions with different functional
forms and then to determine which one shows a better performance, we have considered the
following aperture functions:
\begin{eqnarray}
 w_G(x) & = & \frac{1}{\sqrt{\pi w_G^2}}\ e^{-(x - x_{c,G})^2/w_G^2} , \\
 w_L(x) & = & \frac{1}{\pi w_L} \frac{1}{1 + (x - x_{c,L})^2/w_L^2} , \\
 w_s(x) & = & \frac{1}{\pi w_s}\ {\rm sinc} \left[ (x - x_{c,s})/w_s \right] .
\end{eqnarray}
In particular, for a comparison on the same footing conditions, we have considered $x_c=0$ in the
three cases, while $w$ has been chosen in a way that the three functions have the same FWHM.
In this latter regard, we have fixed the width of the Gaussian aperture function to be
$w_G = 26.5$~$\mu$m ($0.5$ in reduced units, in terms of $x_0$), which renders a
${\rm FWHM} = 44.13$~$\mu$m and the following width values for the two other aperture functions:
\begin{eqnarray}
 w_L & = & w_G \sqrt{\ln 2} \approx 22.06\ \mu{\rm m} , \\
 w_s & \approx & \frac{w_G \sqrt{\ln 2}}{u_+} \approx 11.64\ \mu{\rm m} .
\end{eqnarray}
where
\begin{equation}
 u_+ = \frac{\pi}{2} - 4 + \sqrt{16 + 4\pi - \pi^2} \approx 1.895
\end{equation}
is the (positive) solution for ${\rm sinc}\ u$ arising from the Bh\={a}skara I's sine
approximation formula \cite{sanz:PhysScr:2022},
\begin{equation}
 {\rm sinc}\ u \approx \frac{16 (\pi - u)}{5\pi^2 - 4u (\pi - u)} ,
\end{equation}
for the equation ${\rm sinc}\ u = 0.5$, with $u = (x - x_{c,s})/w_s$.
It is worth mentioning that we have considered the same FWHM for the above three aperture
functions instead of in their Fourier spectrum, because the truncation, in the end, is going
to manifest along the spatial transverse section of the beam.
Thus, regardless of the spectrum associated with the above aperture functions, if we set a
given width for all of them, we will control the spatial limitation of the beam.
However, if we choose instead to set the same FWHM for the corresponding Fourier spectra,
we may end up with aperture functions with very different (non comparable) widths.

\begin{figure}[!t]
	\centering
	\includegraphics[width=\textwidth]{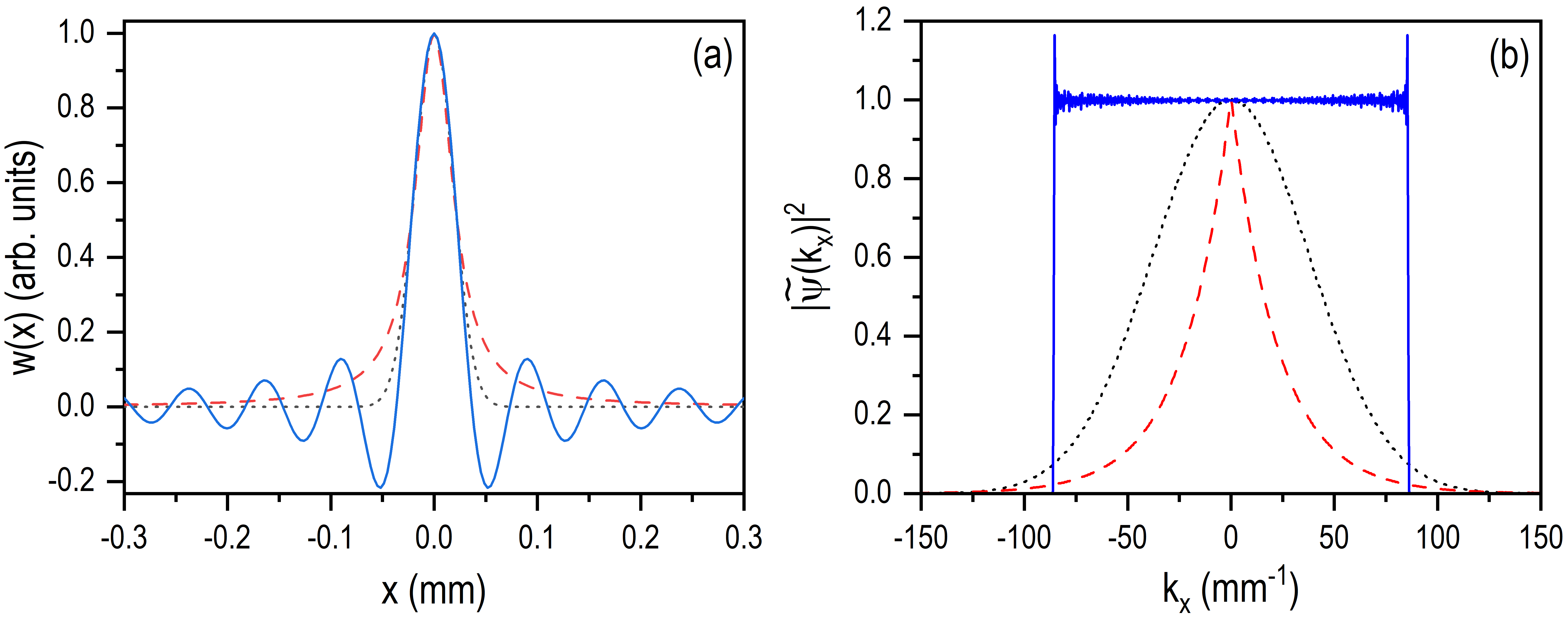}
	\caption{\label{fig1}
		(a) Spatial dependence of the three aperture functions considered here, all with the
		same FWHM, $44.13$~$\mu$m.
		The Gaussian aperture function is represented with black dotted line, the Lorentzian
		aperture function with red dashed line, and the sinc-type aperture function with blue
		solid line.
		For a better comparison, the three functions are normalized to the unity.
		(b) Transverse power spectrum, $|\tilde{\psi}(k_x)|^2$, of the input convoluted wave
		functions arising after applying the convolution with the aperture functions displayed
		in part~(a) [type and color lines are in correspondence with the respective figures
		in (a)].
		Again, for a better comparison, the three spectra have been normalized to the unity
		at their center.}
\end{figure}

The three aperture functions are shown in Fig.~\ref{fig1}a, all with the same FWHM,
$44.13$~$\mu$m, which is nearly half the FWHM of the leading maximum of the ideal
(infinite energy) Airy beam considered, $96.13$~$\mu$m.
For a better comparison, the three functions are plotted without the constant prefactor,
so that their maxima coincide with 1, while they asymptotically approach 0 at large $x$.
We notice that, although the three cases present nearly the same trend around $x_c$,
the spatial extent of the sinc-type function is larger than in the other two cases.
Nonetheless, if we inspect the transverse power spectrum, $|\tilde{\psi}(k_x)|^2$,
corresponding to the initial ans\"atze constructed with these aperture functions
(see Fig.~\ref{fig1}b), we notice that the three cases
essentially span the same range of momenta, even though the profile of the respective
spectra looks very different (again, for a better comparison, the three spectra have
been normalized to the unity at their center).
Given that the power spectrum of an ideal Airy beam is flat, with an infinite extension,
we note that, while the Gaussian and Lorentzian aperture functions are going to
introduce a gradually decreasing of the amplitude components involved in the Airy
beam, the sinc-type aperture function, on the contrary, is going to essentially
preserve the same amplitude for the spectral range spanned.
In the current case, as it can be seen in Fig.~\ref{fig1}b, this spectral range is
$\Delta k_x = 2/w_s \approx 171.76$~mm$^{-1}$, with all components, low and large
spatial frequencies, essentially contributing the same.
This is in sharp contrast with the other two aperture functions, where the low spatial
frequencies will play a more prominent role than the large ones (more remarkable in
the Lorentzian case than in the Gaussian one).


\begin{figure}
 \centering
 \includegraphics[width=\textwidth]{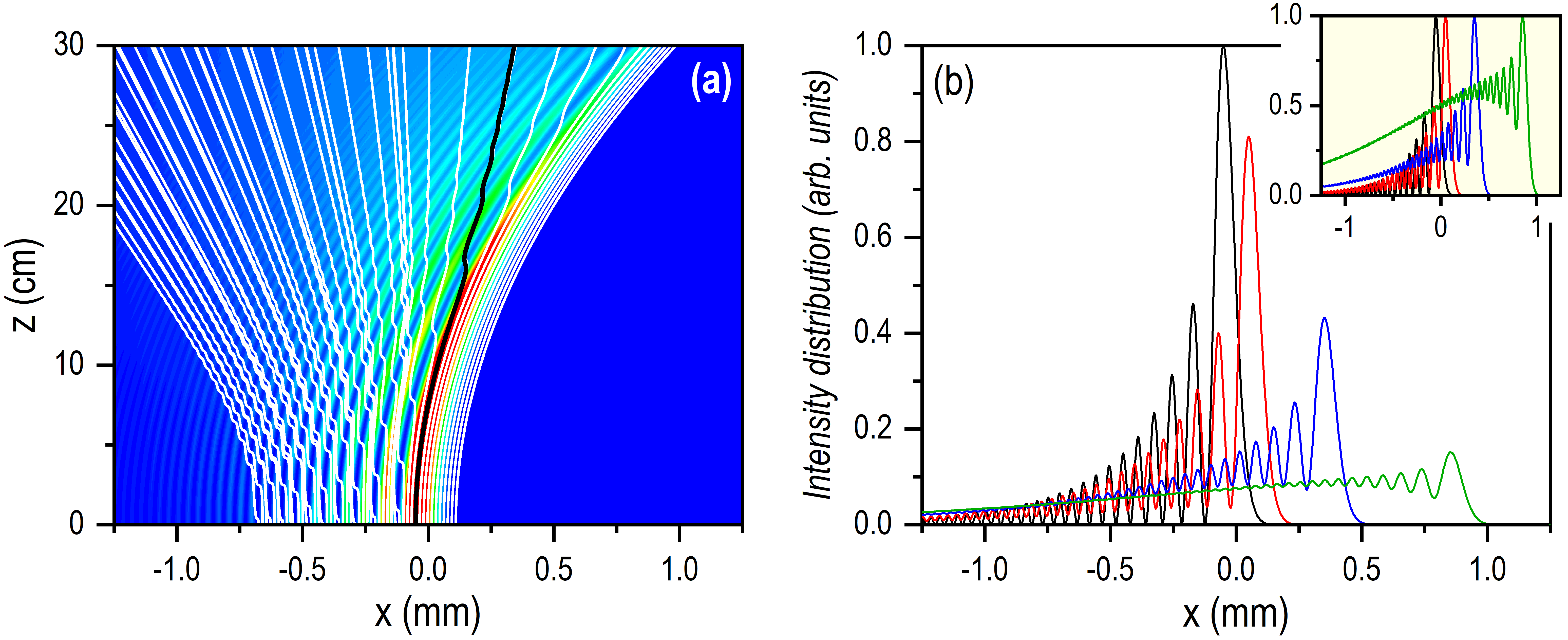}
 \caption{\label{fig2}
  (a) Flux trajectories (white solid lines) associated with an Airy beam convoluted with a
  Gaussian aperture function.
  The initial conditions cover homogeneous the range spanned by the first foremost 10 maxima
  (including the leading one).
  The thick black line denotes a trajectory with its initial condition at the same position
  where the main maximum lies.
  (b) Profiles of the intensity distribution at: $z=0$ (black solid line), $z=10$~cm (red solid line),
  $z=20$~cm (blue solid line), and $z=30$~cm (green solid line).
  The curves have been normalized to unity taking into account the maximum value of the input beam
  (black solid line).
  In the inset, the same curves but each normalized to unity with respect to their corresponding maxima.}
\end{figure}

\section{Results}
\label{sec3}


\subsection{Numerical dynamical simulations}
\label{sec31}

\begin{figure}
 \centering
 \includegraphics[width=\textwidth]{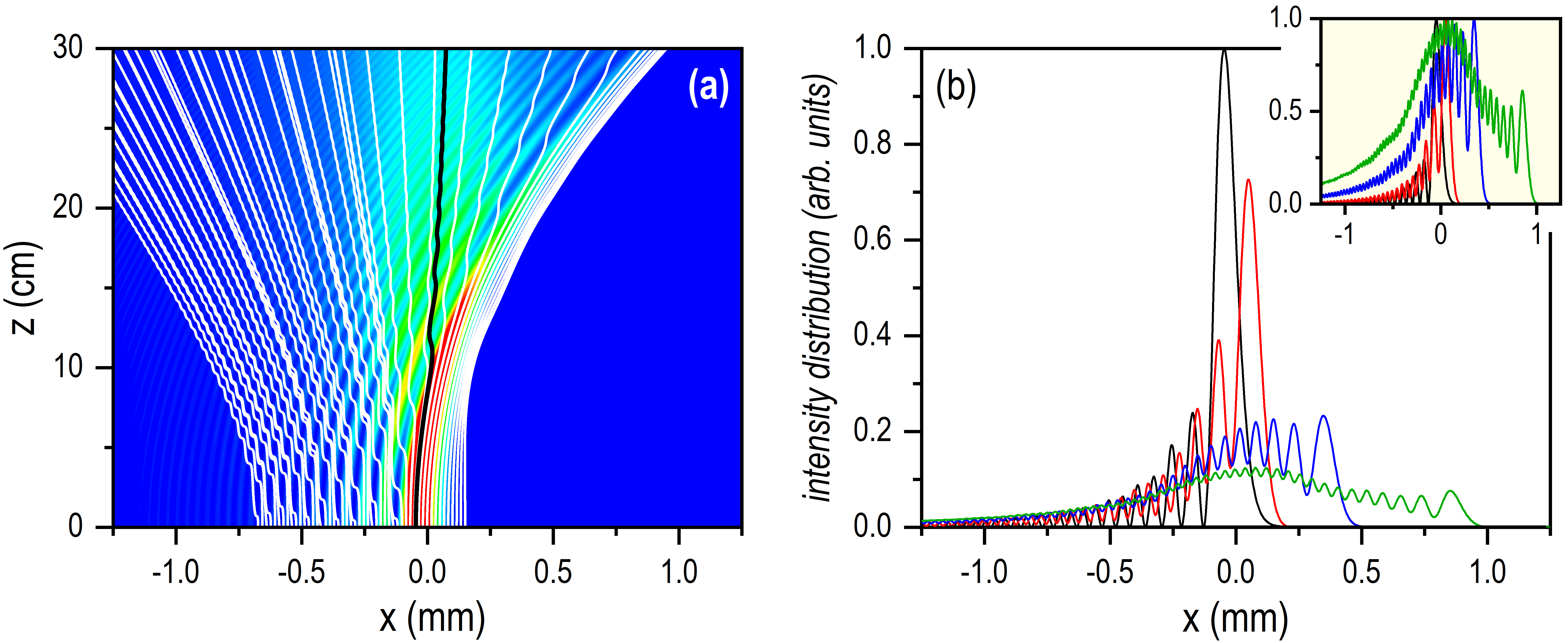}
 \caption{\label{fig3}
  (a) Flux trajectories (white solid lines) associated with an Airy beam convoluted with a
Lorentzian aperture function.
The initial conditions cover homogeneous the range spanned by the first foremost 10 maxima
(including the leading one).
The thick black line denotes a trajectory with its initial condition at the same position
where the main maximum lies.
(b) Profiles of the intensity distribution at: $z=0$ (black solid line), $z=10$~cm (red solid line),
$z=20$~cm (blue solid line), and $z=30$~cm (green solid line).
The curves have been normalized to unity taking into account the maximum value of the input beam
(black solid line).
In the inset, the same curves but each normalized to unity with respect to their corresponding maxima.}
\end{figure}

\begin{figure}
 \centering
 \includegraphics[width=\textwidth]{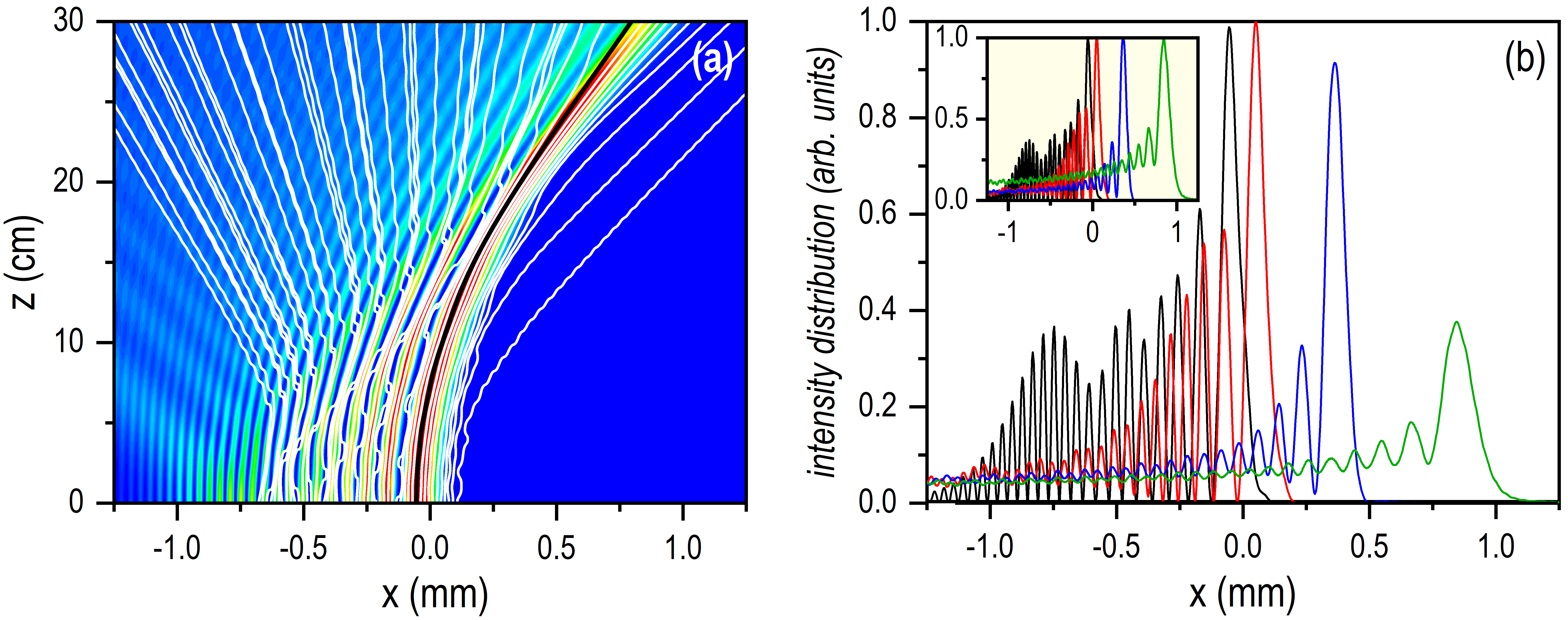}
 \caption{\label{fig4}
  (a) Flux trajectories (white solid lines) associated with an Airy beam convoluted with a
sinc-type aperture function.
The initial conditions cover homogeneous the range spanned by the first foremost 10 maxima
(including the leading one).
The thick black line denotes a trajectory with its initial condition at the same position
where the main maximum lies.
(b) Profiles of the intensity distribution at: $z=0$ (black solid line), $z=10$~cm (red solid line),
$z=20$~cm (blue solid line), and $z=30$~cm (green solid line).
The curves have been normalized to unity taking into account the maximum value of the beam at $z=10$~cm
(red solid line).
In the inset, the same curves but each normalized to unity with respect to their corresponding maxima.}
\end{figure}

To check the feasibility of the aperture functions here proposed, we have conducted a series
of numerical simulations due to the non-analiticity of the propagation process in the three
cases.
As it is mentioned above, the parameter values considered are analogous to those used in the
original experiments \cite{christodoulides:PRL:2007}, which enables a more direct comparison
with such experiments.
Accordingly, in Figs.~\ref{fig2}, \ref{fig3}, and \ref{fig4} we show both an overall view of
the propagation process in terms of trajectories superimposed on the corresponding density
plots of the intensity distribution (see part~a in these figures), and a series of snapshots
of the intensity distribution at given values of $z$, which allow us to elucidate in a
qualitative manner how fast the properties of ideal Airy beams are gradually lost (see
part~b in the figures).

In Fig.~\ref{fig2}a we observe the propagation for a beam obtained with the Gaussian
aperture function.
As it is seen from the density plot, the beam behaves similarly to an ideal Airy beam for
a relatively long distance, with a well-defined main leading maximum surviving up to nearly
the largest $z$-value considered ($z=30$~cm).
The snapshots displayed in Fig.~\ref{fig2}b show clear evidence of such a behavior,
particularly in the inset in this figure, where we have renormalized the three distribution
for a better comparison.
Nonetheless, unlike an ideal Airy beam, there is a progressive decrease in the intensity as
well as a loss of the characteristics oscillations that form the tail of the beam.
To understand this behavior, we resort to the trajectory description (see Fig.~\ref{fig2}a).
Accordingly, we note that, as it was mentioned in \cite{sanz:JOSAA:2022}, the limited
extension of the tail releases the beam from the infinite energy influx coming from the back.
Consequently, not only the beam propagates forward, but it can also get diffracted backwards,
which is what the trajectories (white solid lines) clearly show.
Actually, we can observe that, if we choose an initial condition at the position of the
maximum of the main peak of the initial beam, the corresponding trajectory (black solid
line) start deviating from a parabolic-type curve for distances beyond $z=15$~cm,
approximately.
In other words, typical Airy beam properties are warranted for about $15$~cm; afterwards,
the beam starts getting distorted.

The results for the Lorentzian aperture function are shown in Fig.~\ref{fig3}.
In this case, we observe a faster degradation, at shorter distances (around $10$~cm).
In Fig.~\ref{fig3}a the density plot reveals a large transfer of flux from the main peak
to the rear secondary maxima, although with the seemingly net effect of getting stacked
around $x = 0$~mm.
If we inspect Fig.~\ref{fig3}b, effectively we clearly notice an accumulation of intensity
around $x = 0$~mm (see also inset), decreasing towards larger values of $x$.
From the trajectory description, we find an overall picture that resembles that of the
Gaussian aperture, but with the subtle difference that now the trajectory starting at the
position of the main maximum remains nearly around the same position, without following
the foremost ensemble of trajectories that still show a parabolic-type shape.

The two previous cases can be understood in the light of the strong filtering produced
by the aperture functions considered in each case, which remove dramatically the high
frequencies from the spectrum (see Fig.~\ref{fig1}a).
In sharp contrast, the case of the sinc-type aperture function, represented in
Fig.~\ref{fig4}, shows a very different picture.
First, except from some intensity bursts propagating backwards, the density plot shows that
all maxima, both the main peak and the secondary maxima, exhibit the typical parabolic shape
expected for an Airy beam.
The results in Fig.~\ref{fig4}b clearly indicate that, effectively, the beam behaves like an
ideal Airy beam for longer distances.
On the other hand, the behavior of the trajectories also reveals that there is a larger
survival of the parabolic shape for longer distances.
Unlike the two previous cases, note that here the trajectories show a back and forth
displacement, which keeps confined along parabolic-like channels for longer distances,
until the backward energy influx reduces and they can escape (backwards).
This behavior makes that a large portion of the foremost trajectories will keep showing
their parabolic shape for rather long $z$-distances.
Indeed, note that the trajectory started at the position of maximum of the main peak is
nearly parabolic for the full propagation.


\subsection{Dynamics of the main maximum}
\label{sec32}

Apart from a local description of the propagation process, the trajectories also render
interesting information that cannot be directly obtained from the density plots.
In particular, to further understand physically the performance of the three aperture
functions, we can investigate the local dynamics associated with a piece of the beam
without perturbing its propagation.
As it is inferred from \cite{sanz:JPA:2011}, the intensity content within two any initial
conditions will remain constant along $z$ provided we propagate those conditions along the
corresponding trajectories.
That is, if we consider two initial conditions delimiting a given intensity maximum (for
instance, setting such initial conditions near the adjacent nodes, but not on top of them),
we can determine at any subsequent $z$-value how the corresponding intensity distributes
and within which boundaries.

\begin{figure}
	\centering
	\includegraphics[width=\textwidth]{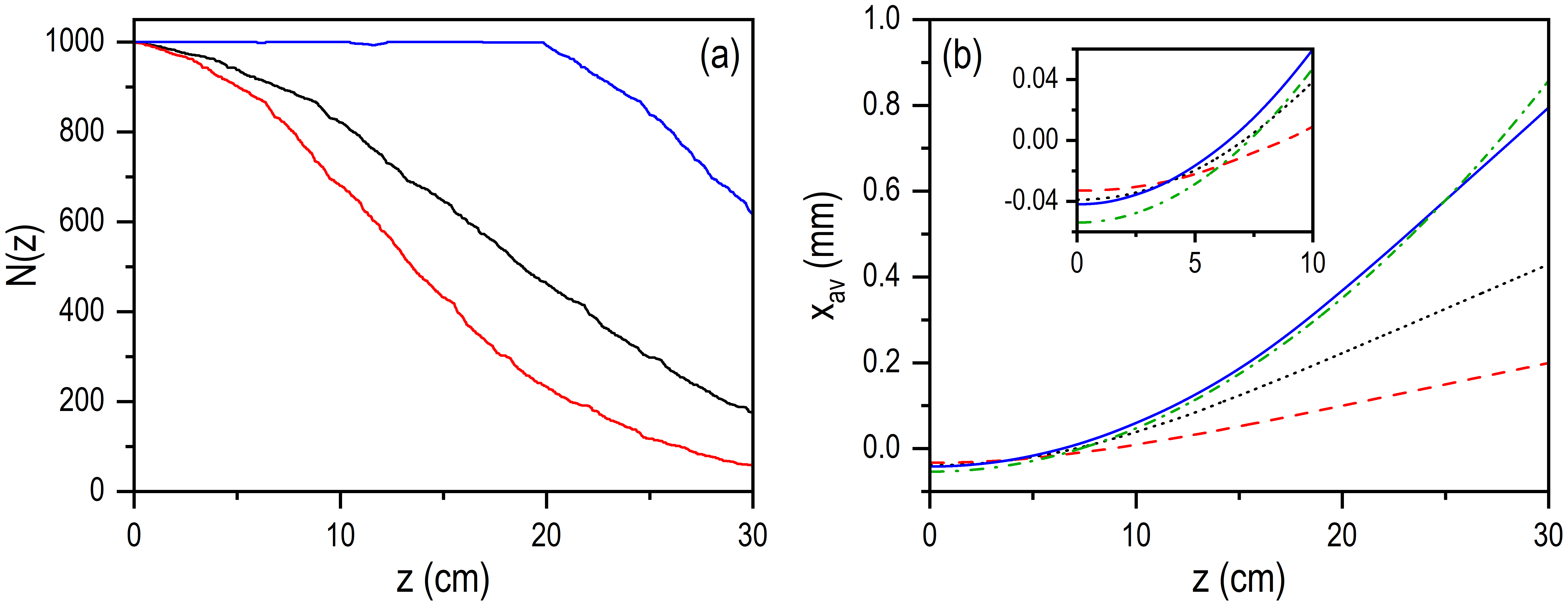}
	\caption{\label{fig5}
		(a) Escape rate, $N(z)$, as a function of $z$ for the three aperture functions here
		considered.
		(b) Average position, $x_{\rm av}(z)$, over a sampling of 1,000 flux trajectories randomly
		distributed, at $z=0$, within the area covered by the leading maximum for the three
		aperture functions.
		The distribution of initial conditions have been taken according to a biGaussian
		distribution (see Appendix~\ref{app}).
		To compare with, the $z$-dependence of the leading maximum of an ideal (infinite energy)
		Airy beam is also shown, represented with green dashed line.}
\end{figure}

Accordingly, we have considered pairs of initial conditions between which we collect most of
the intensity associated with the main peaks in the three cases (the ranges are given in
Table~\ref{tab1} in Appendix~\ref{app}).
Then, in order to understand how fast the corresponding beams deviate from an ideal Airy
beam, we have computed two quantities.
On the one hand, we have computed the escape rate, $N(z)$, which measures, along
$z$, the intensity confined within two boundaries given by two parabolic trajectories.
It is clear that the better the preservation of the ideal Airy beam properties the larger
the value of $N(z)$ for larger and larger $z$-values.
This quantity is plotted in Fig.~\ref{fig5}a.
As it can be seen, the decay of $N(z)$ is relatively fast for the Gaussian and the
Lorentzian aperture functions, although the falloff for the latter is still slightly
larger.
While the energy (proportional to the number of trajectories) decays to about a half
at $z \approx 18.8$~cm for the Gaussian aperture function, in the Lorentizan case the
same takes place at $z \approx 13.5$~cm.
This corresponds to the degradation of the intensity distribution profile observed in
Figs.~\ref{fig2}a and \ref{fig3}a, respectively.
On the contrary, for the sinc-type aperture function, $N(z)$ remains nearly constant
until $z \approx 19.8$~cm and, extrapolating data, a decrease to about $50\%$ is only
expected for larger distances ($z \approx 33.5$~cm), which is in correspondence with
the behavior observed in Fig.~\ref{fig4}a.

On the other hand, taking into account that those initial conditions have been randomly
distributed according to the intensity distribution, their average, $x_{av}$, can provide
an objective measure of the self-accelerated propagation of only this beam.
Again, the better the preservation of the ideal behavior, the closer this average will be
to a parabola.
This quantity is represented in Fig.~\ref{fig5}b (see solid lines), where we have also
included the trajectories associated with the initial condition taken at the position of the
maximum of the respective main peaks (dashed lines).
To compare with, we have also included the trajectory of an ideal Airy beam with its initial
position at the $x$-value where the main peak of such a beam reaches its maximum (see green
dotted line).
As we can see, the trajectories corresponding to the maximum in the Gaussian and Lorentzian
cases start deviating from the average value at short and medium distances, respectively, in
agreement with what we already saw in Figs.~\ref{fig2} and \ref{fig3}.
More specifically, such a deviation starts being remarkable at about $20$~cm for the Gaussian
aperture function, and about $10$~cm for the Lorentzian one.
These distances can be used as distinctive objective criterion to determine when the
performance of the corresponding aperture function becomes poor.
In therms of the average, $x_{av}$, this loss of efficiency is indicated by the transition
from a nearly parabolic behavior, at the initial stages of the propagation, to a seemingly
linear behavior at larger distances, which is produced by the gradual drain of trajectories
(flux) backwards.

In the case of the sinc-type aperture function, again we observe a very different
behavior, since both the average and the trajectory of the maximum basically propagate
parallel one another.
More interestingly, these two quantities overlap in a good approximation the trajectory of
the maximum for an ideal Airy beam; only at large $z$-values we start observing a
discrepancy, which is a signature of the loss of the ideal behavior.

	
\section{Discussion}
\label{sec4}

We have introduced a trajectory-based methodology to investigate the effect of
limiting the energy content of Airy beams by means of aperture functions with
different shapes, and hence to elucidate which functional form better preserves
the properties of an ideal Airy beam at the input.
Thus, we have combined conventional propagation simulations, to determine the intensity
distribution at difference distances from the input plane, with the above mentioned
non-conventional approach based on an on-the-fly computation of the transverse energy
flux, which provides us with information at a more local level.
The transverse energy flux, described in terms of sets of trajectories,
enables a description and analysis of the propagation process on a single-event basis,
making more intuitive the properties of Airy beams (shape invariance and self-accelerated
propagation) at the same time that it is also possible to perform some additional
quantitative analyses of specific features.

From our analysis, we conclude that, among the three types of aperture functions
investigated, sinc-type ones show a better
performance, because they basically introduce a cut in the beam spectrum, without
modifying the amplitude of its components, unlike other types of aperture functions,
such as Gaussian or Lorentzian profiles.
The trajectories show a very interesting picture in this regard.
On the one hand, we note that the trajectories associated with the Gaussian or the
Lorentzian aperture functions make evident a monotonous energy dissipation backwards.
This continuous flow produces a rather fast lost of fringe visibility of the intensity
distribution, as it is manifested by the corresponding escape rates, and hence the
also fast loss of the typical properties of an Airy beam at relatively close distances
from the input plane.
In particular, it has been noticed that the self-accelerated displacement that should
undergo the average position of the energy accumulated in the main maximum is gradually
lost.

On the other, in the case of the sinc-type aperture function, the fact that all spatial
spectral components play have nearly the same amplitude provokes that the presence of
a wiggling displacement of the trajectories and hence a back-and-forth energy flow
between adjacent intensity maxima.
This oscillatory behavior enables the preservation of the typical Airy beam features
for much longer distances, even though the gradual degradation observed in the pattern
along them.
Furthermore, unlike the two previous cases, here it has been shown that the average
position of the energy accumulated in the main maximum follows very nicely the behavior
expected for an ideal Airy beam.

In sum, the present analysis is aimed at providing new insights into the behavior of
finite-energy Airy beams, and therefore also expected to contribute to the design and
development of novel applications exploiting finite-energy Airy beams.
Moreover, it is also worth noting that, by virtue of the tight analogy between the
paraxial Helmholtz equation and the Schr\"odinger equation, the results shown and
discussed here are also directly transferable to the generation and exploitation of
Airy beams with nonzero mass particles, as it is the case of electron Airy beams,
first produced in 2013 by Voloch-Bloch {\it et al.}\ \cite{voloch:Nature:2013}.


\appendix
\section{Numerical and data analysis methods}
\label{app}

The simulations conducted in this work to obtain the beam propagation and the corresponding
trajectories, have been carried out following the same procedure formerly used in
\cite{sanz:JOSAA:2022}, based on the split-step for beam propagation and a spectral
recast of the equation of motion for the trajectories \cite{sanz-bk-2}.
For the trajectories shown in Figs.~\ref{fig2}, \ref{fig3}, and \ref{fig4}, we have
considered an evenly-spaced distribution of 50 initial conditions, taken from a range
that covers the first 10 intensity maxima of the input beam (the leading maximum and
the nine subsequent secondary maxima).

Concerning the random distribution required to compute the quantities shown in
Fig.~\ref{fig5}, first we have fitted the main maximum of the input intensity distributions
(see black line curves in Figs.~\ref{fig2}, \ref{fig3}, and \ref{fig4}) to the biGaussian
distribution,
\begin{equation}
 f(x) = \left\{
 \begin{array}{lcc}
  \exp [-(x-x_0)^2/2 \sigma_l^2] , & \qquad & x < x_0 \\
  \exp [-(x-x_0)^2/2 \sigma_r^2] , & \qquad & x \ge x_0 \\
 \end{array}
 \right. ,
\end{equation}
which captures fairly well the asymmetry of such a maximum.
The values for the parameters $x_0$, $\sigma_l$, and $\sigma_r$ obtained from the
corresponding fitting procedures for the three cases are given in Table~\ref{tab1}.

\begin{table}[t]
 \caption{\label{tab1} Parameters used in the fitting of the main maximum to a biGaussian distribution.}
 \centering
 \begin{tabular}{ccccc}
 \hline
 aperture function & $x_c$ (mm) & $w_1$ (mm) & $w_2$ (mm) & i.c.\ range (mm) \\
 \hline \hline
 Gaussian   & $-0.05688$ & $0.02647$ & $0.04844$ & $(-0.123,0.115)$ \\
 Lorentzian & $-0.05363$ & $0.02887$ & $0.05412$ & $(-0.128,0.161)$ \\
 sinc-type  & $-0.06016$ & $0.02341$ & $0.04536$ & $(-0.124,0.112)$ \\
 \hline
\end{tabular}
\end{table}

Regarding the specific random choice, first, we select randomly from a $[0,1)$ uniform
distribution a number.
If this number is smaller than the ratio $\sigma_l/(\sigma_l + \sigma_r)$, then we associate
the initial condition to the left side of the biGaussian distribution (i.e., $x < x_0$);
otherwise, the initial condition is associated with the right side.
A number is then picked up randomly following a Gaussian distribution centered at $x_0$ and
with a width $\sigma_l$, if the previous choice laid on the left side of the distribution,
or $\sigma_r$, if the choice led to the right side.
It might happen that the number obtained from this procedure will lay on the opposite side
where it is supposed to be; in such a case, the algorithm re-start the choice of the side
and repeats the process.


\section*{Funding}
Agencia Estatal de Investigaci\'on (PID2019-104268GB-C21, PID2021-127781NB-I00, RED2022-134391-T).

\section*{Disclosures}
The authors declare no conflicts of interest.

\section*{Data Availability}
Data underlying the results presented in this paper are not publicly available,
but may be obtained from the authors upon reasonable request.





\end{document}